\documentclass[twocolumn,showpacs,preprintnumbers,amsmath,amssymb,superscript address, revsymb,prb]{revtex4}

\usepackage[dvips]{graphicx}
\usepackage{mathptmx}
\usepackage{dcolumn}

\newcommand{\etal}{\emph{et al.}}
\newcommand{\defect}[1]{$\{I,{\rm H}_{#1}\}$}
\newcolumntype{.}{D{.}{\cdot}{3.10}}

\begin{document}

\title{Hydrogen/silicon complexes in silicon from computational searches}


\author{Andrew J. Morris\footnote{Email: ajm255@cam.ac.uk.}}
\affiliation{Theory of Condensed Matter Group, Cavendish Laboratory,
University of Cambridge, J. J. Thomson Avenue, Cambridge CB3 0HE,
United Kingdom.}

\author{Chris J. Pickard} \affiliation{Scottish Universities Physics
Alliance, School of Physics and Astronomy, University of St Andrews,
North Haugh, St Andrews KY16 9SS, United Kingdom.}

\author{R. J. Needs} \affiliation{Theory of Condensed Matter Group,
Cavendish Laboratory, University of Cambridge, J. J. Thomson Avenue,
Cambridge CB3 0HE, United Kingdom.}

\date{\today{}}

\begin{abstract} 
Defects in crystalline silicon consisting of a silicon
self-interstitial atom and one, two, three, or four hydrogen atoms are
studied within density-functional theory (DFT).  We search for
low-energy defects by starting from an ensemble of structures in which
the atomic positions in the defect region have been randomized.  We
then relax each structure to a minimum in the energy.  We find a new
defect consisting of a self-interstitial and one hydrogen atom
(denoted by \{$I$,H\}) which has a higher symmetry and a lower energy
than previously reported structures.  We recover the \{$I$,H$_2$\} defect
found in previous studies and confirm that it is the most stable such
defect.  Our best  \{$I$,H$_3$\} defect has a slightly different structure
and lower energy than the one previously reported, and our lowest
energy  \{$I$,H$_4$\} defect is different to those of previous
studies. 
\end{abstract}

\maketitle

\section{Introduction} 

Hydrogen is a very common impurity in semiconductors whose roles in
silicon include passivating surfaces and
defects.\cite{VdWalle:ARMR:2006}
Much is known about the vacancy in silicon but rather little
understanding of self-interstitials has been gleaned from experiments,
so there is considerable reliance on theoretical work.
Self-interstitials are common in silicon and they are expected to
react with impurities to form defect complexes.
Mobile hydrogen atoms are expected to bind strongly to
self-interstitial defects in silicon, and hydrogen-silicon complexes
have been detected in experiments.\cite{Budde:PRB:1998}

Silicon self-interstitials are readily formed during device
manufacture and bombardment with electrons or ions.
According to DFT calculations, the most stable structure is the
split-$\langle110\rangle$ defect, with the hexagonal interstitial
being slightly higher in energy and the tetrahedral interstitial being
still higher in
energy.\cite{Needs:JPhysCM:1999,Goedecker:PRL:2002,Batista:PRB:2006}
The results of two quantum Monte Carlo calculations are consistent
with these three defects having low
energies.\cite{Leung:PRL:1999,Batista:PRB:2006}
Mukashev \etal\cite{Mukashev:PSSa:1998} attributed the AA12 electron
paramagnetic resonance center to a self-interstitial defect, possibly
a single self-interstitial.
Calculations by Eberlein \etal\cite{Eberlein:PhysicaB:2001} found the
doubly-positively-charged single self-interstitial to be stable at the
tetrahedral site, and that it was broadly consistent with the AA12
defect.

Much of what is known about hydrogen in silicon has been learnt from
studies of vibrational modes which are accessible to infrared
absorption experiments and may also be calculated within
first-principles methods.
Only one hydrogen-silicon complex has so far been firmly identified in
experiments.
Budde \etal\cite{Budde:PRB:1998} identified the silicon
self-interstitial with two hydrogen atoms using Fourier transform
infrared (FTIR) absorption spectroscopy.
The observed properties were found to be in excellent agreement with
the results of DFT calculations.\cite{Budde:PRB:1998}

Throughout this paper we denote the silicon self-interstitial atom by
$I$ and the $n$ hydrogen atoms by ${\rm H}_n$, and the whole defect is
referred to as \defect{n}.
A metastable defect is indicated by an asterisk.  
We also use the notation devised by Gharaibeh
\etal\cite{Gharaibeh:PRB:2001} where $(n)$-$(m)\cdots$ means $n$
hydrogen atoms bonded to one silicon atom and $m$ hydrogen atoms bound
to a neighboring silicon atom, \emph{etc.}
For example, the silicon self-interstitial bonded to two hydrogen
atoms mentioned above is referred to as \defect{2} and the arrangement
of H atoms shown in Fig.~\ref{Fig:I1H2} is described as (1)-(1).

The first calculations we are aware of on the \defect{} defect were by
D\'eak \etal,\cite{Deak:MSE:1989} who reported two possible structures
which, however, are different from those found subsequently by Van de
Walle and Neugebauer.\cite{VdWalle:PRB:1995}
In each structure found by D\'eak \etal\cite{Deak:MSE:1989} the second
nearest neighbour silicon atom to the hydrogen has a dangling bond,
whereas Van de Walle and Neugebauer's lowest energy structure does not
contain dangling bonds.
Budde \etal\cite{Budde:PRB:1998} found \defect{} structures based on
split-$\langle110\rangle$ and split-$\langle100\rangle$
self-interstitials both of $C_{1h}$ symmetry, with the
$\langle110\rangle$ defect being 0.24\,eV lower in energy.
The \defect{} defect based on the split-$\langle110\rangle$ interstitial was
similar to that found by Van de Walle and
Neugebauer\cite{VdWalle:PRB:1995} whereas the
split-$\langle100\rangle$ is similar to the lowest energy
structure of D\'eak \etal\cite{Deak:MSE:1989} Gharaibeh
\etal\cite{Gharaibeh:PRB:2001} found a structure similar to Van de
Walle and Neugebauer's\cite{VdWalle:PRB:1995} lowest energy defect.

D\'eak \etal\cite{Deak:MSE:1989,Deak:PhysicaB:1991} also studied the
\defect{2} defect.
Their lowest energy \defect{2} defect is also based on a
split-$\langle110\rangle$ self-interstitial with its two dangling
bonds terminated by hydrogen atoms.
Further evidence in favour of this structure has been obtained in a
variety of
studies.\cite{VdWalle:PRB:1995,Budde:PRB:1998,Hastings:PhysicaB:1999,Gharaibeh:PRB:2001}
Gharaibeh \etal\cite{Gharaibeh:PRB:2001} also found this to be the
most stable of the \defect{n} family of defects.

Hastings \etal\cite{Hastings:PhysicaB:1999} studied the \defect{3}
defect within Hartree-Fock (HF) theory, finding a structure with two
hydrogen atoms bonded to a silicon and a third
bonded to a neighbouring silicon, \emph{i.e.}, a (2)-(1) configuration
of hydrogen atoms.
They found interstitial silyl (SiH$_3$) to be 0.44\,eV higher in
energy and concluded that it is unlikely to be found in bulk silicon.
More complete DFT calculations by the same
group\cite{Gharaibeh:PRB:2001} found a \defect{3} defect with a
(1)-(1)-(1) configuration.

Hastings \etal\cite{Hastings:PhysicaB:1999} studied the \defect{4}
defect within HF theory, finding a ground-state structure similar to
the \defect{2} defect, but with additional hydrogen atoms bonded to
the nearest-neighbor and second-nearest-neighbour silicon atom to the
self-interstitial, resulting in a (1)-(1)-(1)-(1) configuration.
They also found a metastable (2)-(2) configuration 0.2\,eV higher in
energy and showed that interstitial silane (SiH$_4$) is very unlikely
to form.
A later DFT study by the same group\cite{Gharaibeh:PRB:2001} found the
ground state of \defect{4} to be a (3)-(1) configuration. 

In this paper we present calculations for the defects \defect{i},
$i=1,4$, as found by a ``random structure searching'' approach using
first-principles DFT methods.
We describe the random structure searching scheme in Sec.~\ref{random
structure searching}.
The non-standard Brillouin-zone integration scheme we have used is
described in Sec.~\ref{brillouin zone sampling}, and the details of
our DFT calculations and some convergence tests are reported in
Sec.~\ref{DFT calculations}.
Our results are described in Sec.~\ref{results} and our main
conclusions are summarized and discussed in Sec.~\ref{discussion}.

\section{Computational Approach} 

\subsection{Random Structure Searching}
\label{random structure searching}

``Random structure searching'' has already proven to be a powerful
tool for finding structures of solids under high
pressures.\cite{pickard:PRL:2006:silane,csanyi:PRB:2007:graphite,pickard:NatureMaterials:2007:hydrogen,pickard:JChemPhys:2007:h2o,pickard:PRB:2007:alh3}
The basic algorithm is very simple: we take a population of random
structures and relax them.
This approach is surprisingly successful and its performance for large
systems can be improved by imposing constraints.
The constraints we have typically employed in work on high-pressure
phases are to ($i$) choose the initial positions of a local minimum
and randomly displace the atoms by a small amount, ($ii$) insert
``chemical units'' (for example molecules) at random rather than
atoms, ($iii$) search within structures of a particular symmetry, and
($iv$) constrain the initial positions of some atoms.

Pickard and Needs\cite{pickard:PRL:2006:silane} showed that ``random
structure searching'' can be applied to finding structures of point
defects.
In the current work we have searched for silicon self-interstitial defect structures
using a 32-atom body-centered-cubic unit cell of diamond-structure
silicon.
The initial configurations were generated by removing an atom and its
four nearest neighbours, making a ``hole'' in the crystal, and placing
six silicon atoms at random positions within the hole.
The initial configurations therefore consisted of a region of perfect
crystal and a defect region in which atoms are positioned randomly.
This is an example of a constraint of type ($iv$) mentioned above.
Relaxing the members of an ensemble of such initial configurations
generated the split-$\langle110\rangle$, tetrahedral and hexagonal interstitial configurations, which
various DFT calculations have shown to be lowest in
energy.\cite{Budde:PRB:1998,Needs:JPhysCM:1999,Goedecker:PRL:2002,AlMushadani:PRB:2003,Mattsson:PRB:2008}

We explored four choices of the hole used to
generate the initial configurations: ($a$) remove one silicon atom and
its four nearest neighbours; ($b$) remove one silicon atom; ($c$) do
not remove any silicon atoms and take the center of the hole to lie at
the hexagonal site; ($d$) the same as ($c$) but with the hole at the
tetrahedral site.
The initial configurations were generated by placing the appropriate
atoms randomly within cubic boxes of sizes ranging from 2 to 6\,\AA\
centered on the hole.
Choices ($b$), ($c$), and ($d$) generally led to us finding the
structure we believe to be most stable within roughly 100
configurations, while the larger hole of choice ($a$) was less
successful and sometimes failed to find the ground state within 500
configurations.

\subsection{Brillouin Zone Sampling Scheme}
\label{brillouin zone sampling}

The importance of performing accurate Brillouin zone integrations when
calculating defect formation energies has been emphasised by, among
others, Shim \etal\cite{Shim:PRB:2005}
DFT calculations by Gharaibeh \etal\cite{Gharaibeh:PRB:2001} explored
the convergence of BZ sampling for self-interstitial-hydrogen
complexes in silicon.
Instead of standard Monkhorst-Pack (MP)
sampling\cite{Monkhorst:PRB:1976} we have used the multi-k-point
generalization of the Baldereschi mean-value point
scheme\cite{Baldereschi:PRB:1973} outlined by Rajagopal
\etal\cite{Rajagopal:PRL:1994,Rajagopal:PRB:1995}
Three linearly-independent reciprocal lattice vectors (${\bf
b}_1$,${\bf b}_2$,${\bf b}_3$) are chosen, which need not be
primitive, and a $l \times m \times n$ grid of k-points is defined by
\begin{eqnarray}
{\bf k}_{ijk} = \frac{i{\bf b}_1}{l} + \frac{j{\bf b}_2}{m} +
\frac{k{\bf b}_3}{n} + {\bf k}_{\rm B}(l,m,n),
\end{eqnarray}
where
\begin{eqnarray}
i=0,1,\ldots,l-1; \; \; j=0,1,\ldots,m-1; \; \; k=0,1,\ldots,n-1.
\end{eqnarray}
The standard Baldereschi mean-value
point\cite{Baldereschi:PRB:1973} can be written in terms of the ${\bf b}_i$ 
\begin{eqnarray}
{\bf k}_{\rm B} = \alpha_1{\bf b}_1 + \alpha_2{\bf b}_2 + \alpha_3{\bf
b}_3,
\end{eqnarray}
which defines the $\alpha_i$, and the offset ${\bf k}_{\rm B}(l,m,n)$ for the multi-k-point
scheme\cite{Rajagopal:PRL:1994,Rajagopal:PRB:1995} is
\begin{eqnarray}
{\bf k}_{\rm B}(l,m,n) = \frac{\alpha_1{\bf b}_1}{l} +
\frac{\alpha_2{\bf b}_2}{m} + \frac{\alpha_3{\bf b}_3}{n}.
\end{eqnarray}
This is the natural multi-k-point generalization of Baldereschi's
scheme as it corresponds to sampling the Baldereschi mean-value point
of the supercell obtained by choosing direct (real-space) lattice
vectors ($l{\bf a}_1$,$m{\bf a}_2$,$n{\bf a}_3$), where the ${\bf
a}_i$ are related to the ${\bf b}_i$ by the standard dual
transformation.
We will refer to this as the Multi-B scheme.

\begin{figure}
\includegraphics*[width=8cm]{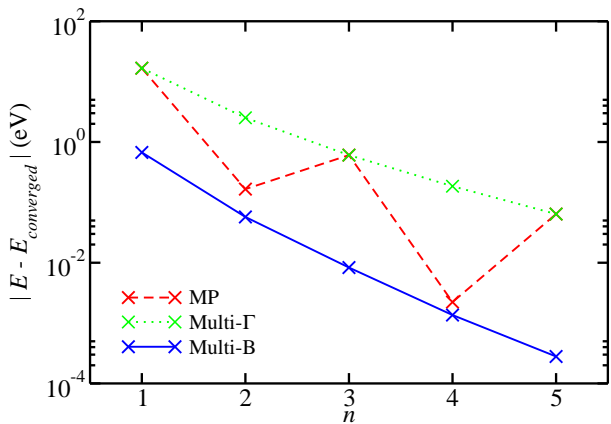}
\caption[]{(Color online) Magnitudes of the energy differences between the converged energy (taken from
a Multi-B calculation with $n=20$) and energies obtained with
different k-point grids for a 2-atom cell of diamond-structure
silicon.  The energies are in eV per cell and the number of points in
each grid is $n^3$.  The standard Monkhorst-Pack grid, labelled MP, is shown in red [dashed line],
Multi-$\Gamma$ denotes grids centered on ${\bf k}$=0 shown in green [dotted line], and Multi-B
denotes the multi-k-point generalization of the Baldereschi scheme is shown in blue [solid line].}
\label{Fig:k-points convergence}
\end{figure}

In Fig.~\ref{Fig:k-points convergence} we report a comparison for
bulk silicon of standard $n\times n \times n$ Monkhorst-Pack grids,
the same grids but centered on ${\bf k}$=0 (Multi-$\Gamma$), and the
Multi-B scheme.
For odd values of $n$ the MP and Multi-$\Gamma$ grids are the same,
although we could of course introduce an appropriate shift of the MP
grid for odd $n$ which gives smooth convergence.
The MP scheme gives smaller errors than the Multi-$\Gamma$ scheme for even $n$.
The Multi-B energies converge smoothly with $n$ and give the smallest errors for each value of $n$.

The cost of a BZ integration is determined not by $n$ but by the
number of symmetry inequivalent k-points in the grid and the most
efficient k-point grid generally depends on the symmetry of the
structure.
Symmetric structures normally have more inequivalent k-points in the
Multi-B grid than in the corresponding MP grid.
During the search stage, symmetry is not imposed and we use all $n^3$
grid points, and therefore the Multi-B scheme is the most efficient.
We have used the Multi-B grid in all searches reported here.

\subsection{Density-functional theory calculations}
\label{DFT calculations}

Our calculations were performed using the Generalized Gradient
Approximation (GGA) density functional of Perdew, Burke and Ernzerhof
(PBE).\cite{Perdew:PRL:1996}
The plane-wave basis-set code {\tt CASTEP}\cite{CASTEP:ZK:2004} was
used with its built-in ultrasoft\cite{Vanderbilt:PRB:1990}
pseudopotentials which include non-linear core
corrections.\cite{Louie:PRB:1982}
All of the results presented here were obtained with
non-spin-polarized calculations.
Some searches and large supercell calculations were repeated allowing
spin-polarization, but no significant changes were found in the energy
differences between structures.

We carried out convergence tests on the formation energy of a
bond-centered hydrogen atom in silicon from interstitial H$_2$ using
256-atom silicon cells.
We chose these systems because they contain very different bonding
arrangements (H$_{\rm bc}$ contains Si-Si and Si-H bonds while the
H$_2$ system contains Si-Si and H-H bonds).
Hence the estimates of convergence of the energy difference between
these two systems should give a reasonable indication of the
convergence of the energy differences between other defects.
The formation energy, $E_{\rm F}$, of a bond-centered H in silicon
from interstitial H$_2$ calculated with $N$-atom silicon cells is
\begin{equation}
E_{\rm F}({\rm H}_{\rm bc},N) = E({\rm H}_{\rm bc},N) - \frac{E({\rm
H}_{2},N)}{2} - \frac{E(N)}{2},
\label{Eqn:EF_bc}
\end{equation}
where $E({\rm X},N)$ is the energy of the X defect in a $N$-atom
silicon cell, and $E(N)$ is the bulk energy for $N$ atoms.

The Fourier transform grid used for wavefunction manipulation was set
to integrate, without aliasing, frequencies twice as high as the
maximum frequency in the basis set.
We checked the convergence of $E_{\rm F}({\rm H}_{\rm bc})$ with
respect to the charge augmentation grid required for the ultrasoft
pseudopotentials, finding it to be converged to within 0.005\,eV at
2.75 times the maximum frequency in the orbital basis set.
We found that $E_{\rm F}({\rm H}_{\rm bc})$ was converged to within
$\pm 0.02$\,eV with a basis set of $E_{\mathrm{PW}}=230$\,eV.

We tested different k-point sampling schemes in the 256-atom cell. 
The values of $E_{\rm F}({\rm H}_{\rm bc})$ for the $n=3$ standard MP
and Multi-B grids agreed to within 0.002\,eV and we therefore
considered these converged. 
The value of $E_{\rm F}({\rm H}_{\rm bc})$ calculated for the 256-atom
cell with the $n=2$ standard MP grid differed from the converged
result by $\sim$0.04\,eV, whereas the error from the Multi-B grid was
six-times smaller.

\section{Calculating the formation energies} 

The  searches were performed using a body-centered-cubic
supercell of a size to contain 32-atoms of bulk silicon.
We used a $n=2$ Multi-B k-point grid, which we estimate gives energy
differences between structures converged to within 0.006\,eV.
However, the 32-atom cell is too small to give highly accurate
geometries and formation energies.
We therefore embedded the most promising structures within 256-atom
body-centered-cubic unit cells and relaxed using a $n=2$ Multi-B grid
until the forces on each atom were less than 0.001\,eV\AA$^{-1}$.

The formation energy of the self-interstitial is defined as
\begin{eqnarray}
E_{\rm F}\left( \left\{I\right\},N\right) = E \left(
\left\{I\right\},N\right) - \frac{N+1}{N}E(N),
\label{Eqn:self-int}
\end{eqnarray}
where $E \left(\left\{I\right\},N\right)$ is the energy of the
self-interstitial cell.

We define the formation energy per hydrogen atom of a system
containing a defect with $n$ hydrogen atoms and $i$ silicon atoms
relative to a system containing $i$ isolated self-interstitial defects
and $n/2$ interstitial hydrogen molecules as
\begin{eqnarray}
E_{\rm F}(\left\{ I_i, {\rm H}_n \right\},N) & = & \frac{ E \left(
\left\{ I_i, {\rm H}_n \right\},N \right) - E \left( \left\{ I_i
\right\},N\right)}{n} + \nonumber \\ && \frac{ E(N) - E \left( {\rm
H}_{2},N \right) }{2}.
\label{Eqn:SiH-int}
\end{eqnarray}
Note that Eq.~(\ref{Eqn:SiH-int}) with $i=0$ and $n=1$ gives
Eq.~(\ref{Eqn:EF_bc}) for the formation energy of bond centered
hydrogen.

\section{Results} 
\label{results}

The formation energies of self-interstitial defects as defined by
Eq.~(\ref{Eqn:self-int}) are given in Table~\ref{Table:self-int}.
In agreement with numerous previous studies (e.g.,
Refs.~\onlinecite{Budde:PRB:1998}, \onlinecite{Needs:JPhysCM:1999},
\onlinecite{Goedecker:PRL:2002}, \onlinecite{AlMushadani:PRB:2003},
and \onlinecite{Mattsson:PRB:2008}) we find the most stable defect to
be the split-$\langle110\rangle$ interstitial.
The hexagonal and tetrahedral interstitials are 0.03\,eV and 0.3\,eV
higher in energy, respectively.
Hence all three self-interstitial defects are candidates for forming a
low energy \defect{n} defect in the presence of hydrogen.
%

\begin{table}[!h]
{\centering \begin{tabular}{lc}
\hline\hline
Defect & $E_{\rm F}\left( \left\{I\right\} \right)$ (eV) \\
\hline
Split-$\langle110\rangle$ & 3.66 \\
Hexagonal                 & 3.69 \\
Tetrahedral               & 3.96 \\
\hline\hline
\end{tabular}\par}
\caption[]{Formation energies for self-interstitial defects in 256-atom cells of silicon, as defined by Eq.~(\ref{Eqn:self-int})
}
\label{Table:self-int}
\end{table}

Our search for hydrogen defects in bulk silicon found the lowest energy
defects known previously, but no new defects.
We found the bond centered H atom, the H$_2$ molecule with the H-H
bond pointing along a $\langle 100 \rangle$ direction, and the
H$_2^{*}$ metastable defect of Chang and Chadi.\cite{Chang:PRB:1989}
The formation energies of hydrogen defects in silicon given in
Table~\ref{Table:EF_H_Si} show the H$_2$ molecule to be the most
stable, as found in previous calculations
\cite{VdWalle:PRL:1988,VdWalle:PRB:1989,VdWalle:PRB:1994}, with the
H$_2^{*}$ defect and bond-centered-hydrogen being respectively
0.10\,eV and 1.04\,eV per H atom higher in energy. 
%

\begin{table}[!h]
{\centering \begin{tabular}{c|c|c|c}
\hline\hline
Defect            & Configuration & $E_{\rm F}$ per H atom (eV) & Degeneracy $d_i$ \\
\hline
H$_2$             & Molecule      &  0.00 & 3 \\
H$^{*}_2$         & (1)-(1)       &  0.10 & - \\
H$_{\mathrm{bc}}$ & Bond Center   &  1.04 & - \\
\hline\hline
\end{tabular}\par}
\caption[]{Formation energies per H atom and degeneracies per atomic
site, $d_i$, of hydrogen defects in silicon. The H$_2$ is the lowest
in energy defect.}
\label{Table:EF_H_Si}
\end{table}

\begin{figure}
\includegraphics*[width=6cm]{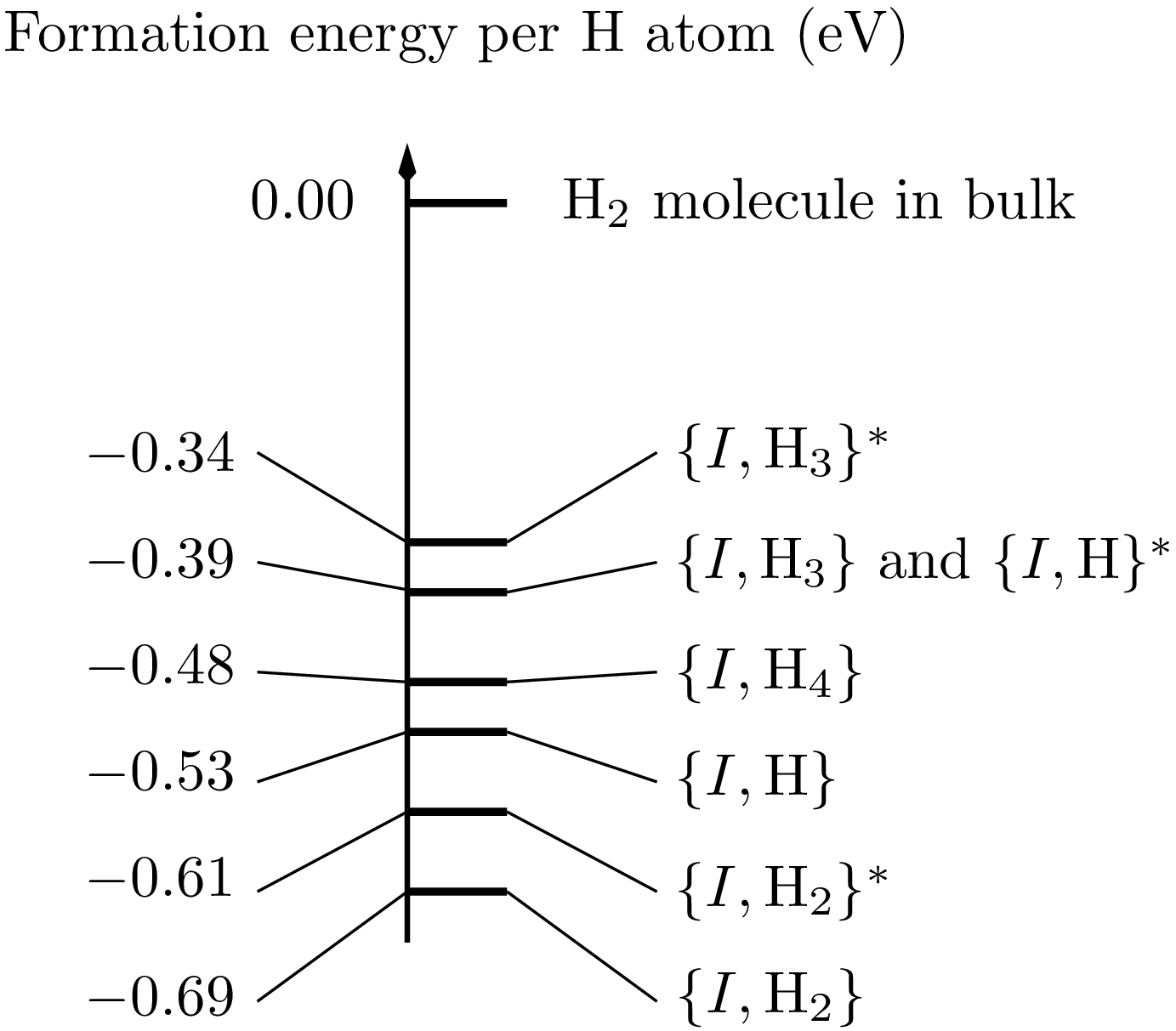}
\caption[]{Formation energies per hydrogen atom using
Eq.~(\ref{Eqn:SiH-int}) for various hydrogen defects in silicon
calculated with respect to silicon containing a self-interstitial
defect and silicon containing hydrogen molecules.}
\label{Fig:MainDiag}
\end{figure}

The formation energies of the hydrogen/silicon self-interstitial
complexes are calculated with reference to a system containing the
most stable self-interstitial (the split-$\langle110\rangle$) and the
most stable hydrogen defect in pure silicon, the H$_2$ molecule.
The data in Table~\ref{Table:EF_H-Si_Si} shows that the \defect{2}
defect has the lowest formation energy, followed by the
\defect{2}$^{*}$.
These data are shown in pictorial form in Fig.~\ref{Fig:MainDiag}.
%

\begin{table}[!h]
{\centering \begin{tabular}{c|c|c|c}
\hline\hline
Defect           &  Configuration  & $E_{\rm F}$ per H atom (eV) & Degeneracy $d_i$ \\
\hline
\defect{}        & (1)             & -0.53 & 4 \\
\defect{}$^{*}$  & (1)             & -0.39 & - \\
\defect{2}       & (1)-(1)         & -0.69 & 12 \\
\defect{2}$^{*}$ & (2)             & -0.61 & - \\
\defect{3}       & (2)-(1)         & -0.39 & 24 \\
\defect{3}$^{*}$ & (1)-(1)-(1)     & -0.34 & - \\
\defect{4}       & (1)-(1)-(1)-(1) & -0.48 & 24 \\
\hline\hline
\end{tabular}\par}
\caption[]{Formation energies per H atom as defined by
Eq.~(\ref{Eqn:SiH-int}) and degeneracies per atomic site, $d_i$, for
various hydrogen/silicon self-interstitial complexes.  The formation
energies are represented pictorially in Fig.~\ref{Fig:MainDiag}}
\label{Table:EF_H-Si_Si}
\end{table}

\begin{figure}
\includegraphics*[width=8cm]{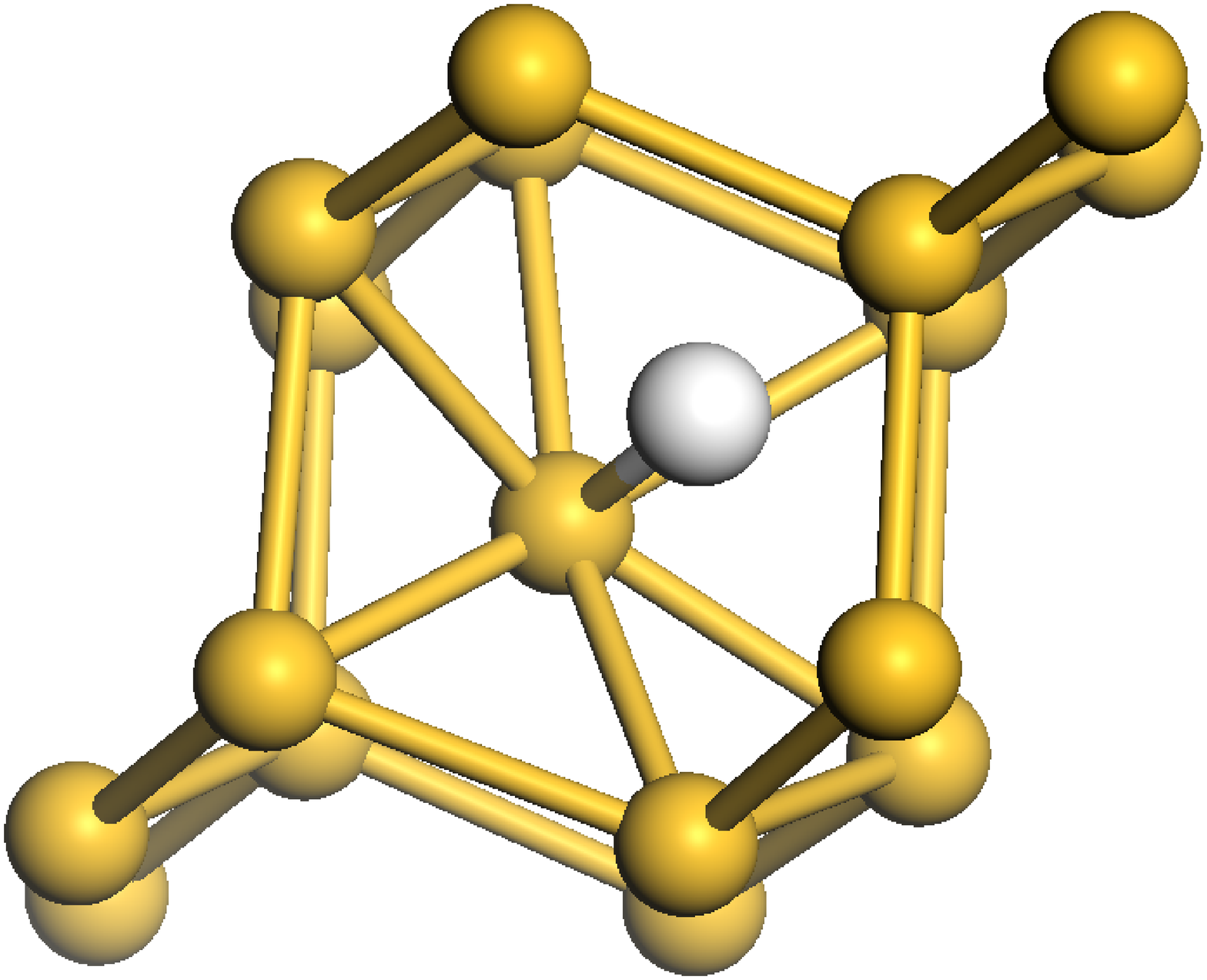}
\includegraphics*[width=8cm]{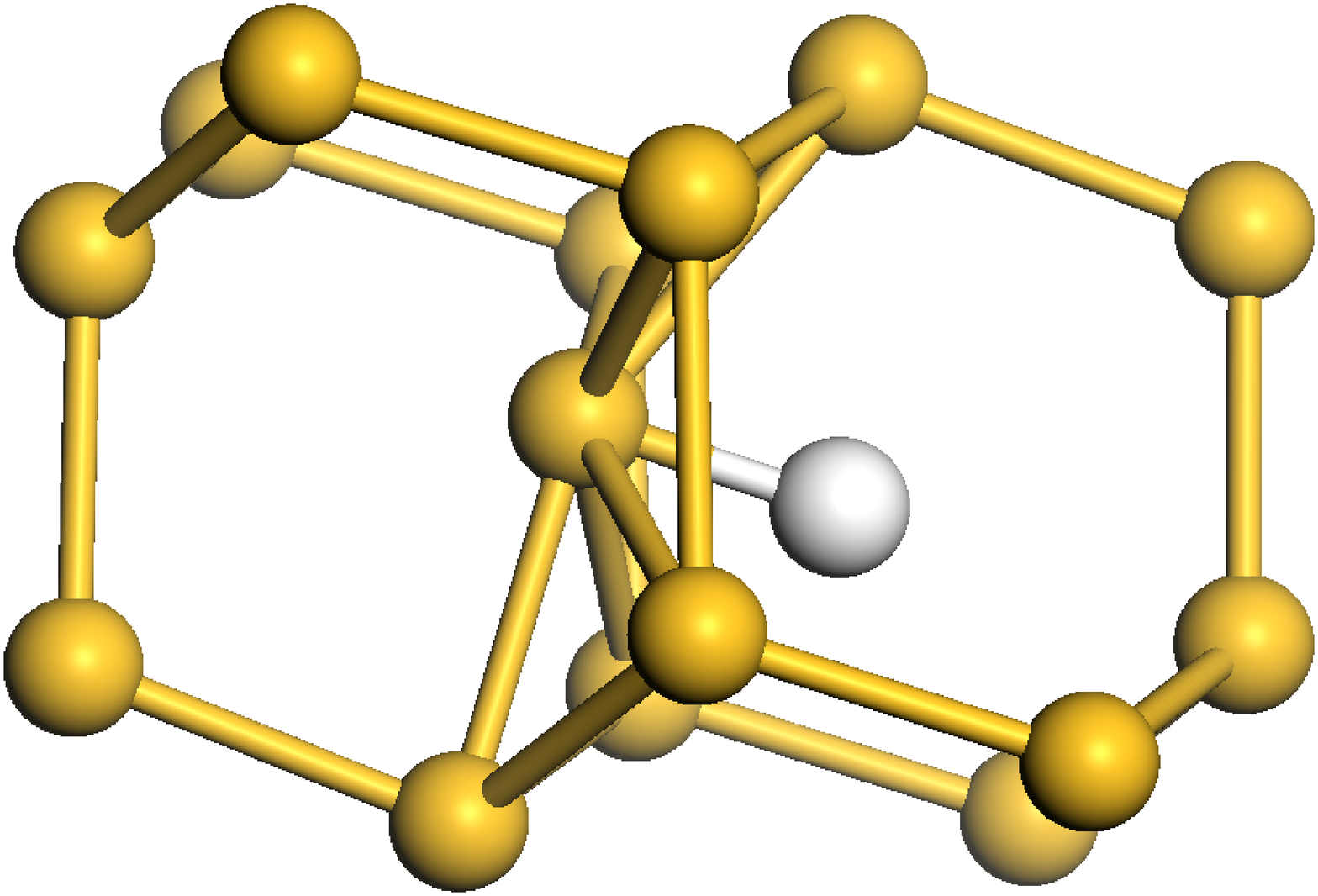}
\caption[]{(Color online) Two projections of the new \defect{} defect
of $C_{3v}$ symmetry found in this work.  Silicon atoms are shown in
yellow (gray) and the hydrogen atom is white.  This defect is based on
a hexagonal self-interstitial rather than the split-$\langle 110
\rangle$  self-interstitial of previous work.  }
\label{Fig:I1H1}
\end{figure}

Our searches for the \defect{} defect found the previously-known
structure\cite{VdWalle:PRB:1995,Gharaibeh:PRB:2001} of $C_s$ symmetry,
but we also found a new lower-energy structure of higher $C_{3v}$
symmetry.
The $C_{3v}$ defect is 0.14\,eV lower in energy and is based on a
hydrogen atom bonded to a hexagonal self-interstitial with the Si-H
bond pointing along a $\langle111\rangle$ direction, rather than the
split-$\langle110\rangle$ interstitial on which the $C_s$ \defect{}
defect is based.

\begin{figure}
\includegraphics*[width=8cm]{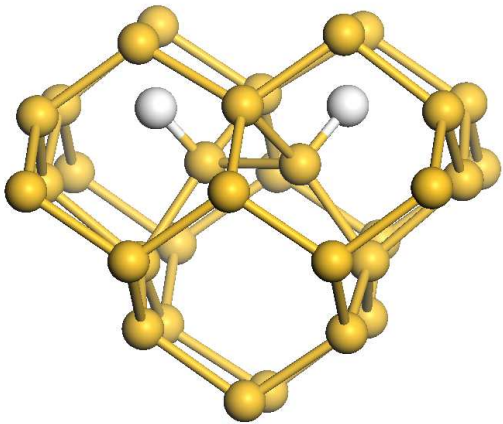}
\caption[]{(Color online) The most stable \defect{2} defect which has
$C_2$ symmetry and was also found in previous
work.\cite{Deak:MSE:1989,Deak:PhysicaB:1991,VdWalle:PRB:1995,Gharaibeh:PRB:2001,Budde:PRB:1998,Hastings:PhysicaB:1999}
Silicon atoms are shown in yellow (gray) and the hydrogen atoms are
white.  The defect is based on a split-$\langle110\rangle$ silicon
interstitial defect, with hydrogen atoms saturating the two dangling
bonds.  }
\label{Fig:I1H2}
\end{figure}

\begin{figure}
\includegraphics*[width=8cm]{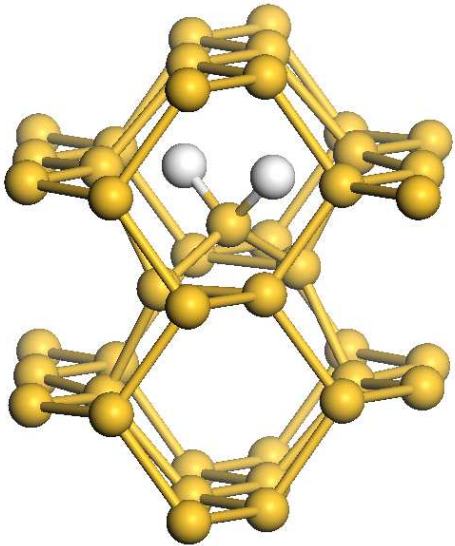}
\caption[]{(Color online) The second most stable \defect{2}$^{*}$
defect which has $C_2$ symmetry and was also found in previous
work.\cite{Deak:MSE:1989} Silicon atoms are shown in yellow (gray) and
the hydrogen atoms are white. The defect is based on a buckled
bond-centered silicon self-interstitial.}
\label{Fig:I1H2star}
\end{figure}

Our searches found the \defect{2} (1)-(1) defect structure reported
previously\cite{Deak:MSE:1989,Deak:PhysicaB:1991,VdWalle:PRB:1995,Gharaibeh:PRB:2001,Budde:PRB:1998,Hastings:PhysicaB:1999}
which is shown in Fig.~\ref{Fig:I1H2} and whose formation energy we
calculated to be -0.69\,eV/H.
The second most stable \defect{2} defect, which has a (2) structure, is
the \defect{2}$^{*}$ defect found previously by D\'eak
\etal\cite{Deak:MSE:1989} and is shown in Fig.~\ref{Fig:I1H2star}.
This defect is based on a buckled bond-centered Si atom with two H
atoms saturating its dangling bonds.
The \defect{2} and \defect{2}$^{*}$ both have $C_2$ symmetry.  
Hastings \etal\cite{Hastings:PhysicaB:1999} found \defect{2}$^{*}$ to be 0.40\,eV
higher in energy than \defect{2} within HF theory.
Later Gharaibeh \etal\cite{Gharaibeh:PRB:2001} found it to be 0.05\,eV
above the ground state. 
Our calculations gave an energy 0.08\,eV/H higher than the \defect{2}
of Fig.~\ref{Fig:I1H2}.

\begin{figure}
\includegraphics*[width=8cm]{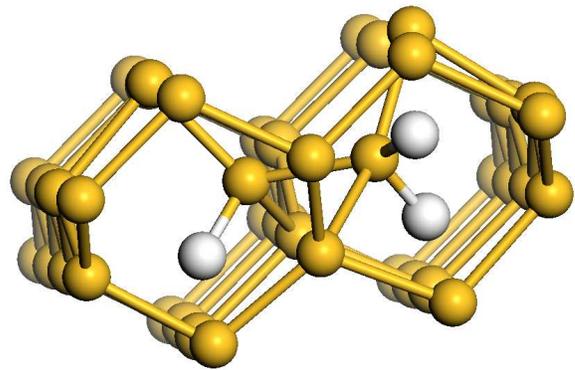}
\caption{(Color online) The new \defect{3} defect of $C_1$ symmetry
found in this work.  Silicon atoms are shown in yellow (gray) and the
hydrogen atoms are white.  The defect is based on a deformed
split-$\langle110\rangle$ self-interstitial.  }
\label{Fig:I1H3}
\end{figure}

The lowest energy \defect{3} defect we found is a (2)-(1)
configuration of $C_1$ symmetry, which is shown in Fig.~\ref{Fig:I1H3}
and has a formation energy of -0.39\,eV/H.
This defect has the same configuration of hydrogen atoms but a
different structure to the one found by Hastings
\etal\cite{Hastings:PhysicaB:1999} using HF theory.
They also found a metastable \defect{3}$^{*}$ structure of $C_1$
symmetry with a (1)-(1)-(1) configuration just 0.1\,eV higher in
energy.
However, more recently this group have used DFT methods and found the
(1)-(1)-(1) configuration to be lower in energy than a (2)-(1)
configuration.\cite{Gharaibeh:PRB:2001} 
We find both the new (2)-(1) and (1)-(1)-(1) configurations mentioned
above, with the new one being 0.05\,eV/H lower in energy than the
(1)-(1)-(1).
%

\begin{figure}
\includegraphics*[width=8cm]{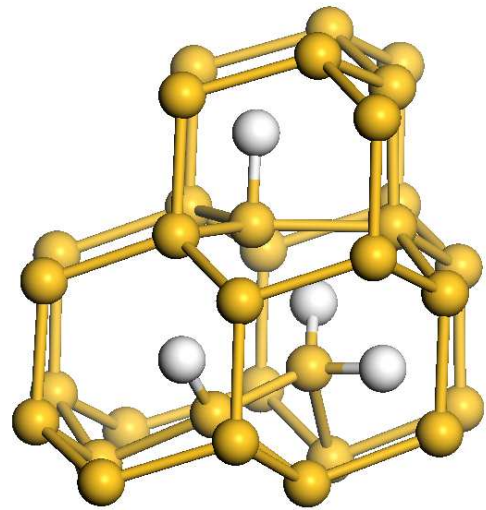}
\caption{(Color online) The new \defect{4} defect of $C_1$ symmetry
found in this work.  Silicon atoms are shown in yellow (gray) and the
hydrogen atoms are white.  It is based on the new \defect{3} (see
Fig.~\ref{Fig:I1H3}) with a defect similar to the H$^{*}_1$ defect of Chadi\cite{Chadi:APL:2003} adjacent
to it on the anti-bonding site.  }
\label{Fig:I1H4}
\end{figure}

The most stable \defect{4} defect we found, shown in Fig.~\ref{Fig:I1H4}, is
made up from the \defect{3} defect of Fig.~\ref{Fig:I1H3} with a
defect similar to the H$^{*}_1$ adjacent
to it at the anti-bonding-type site as shown by
Chadi.\cite{Chadi:APL:2003}
This \defect{4} defect has a (2)-(1)-(1) configuration and $C_1$
symmetry.
Our \defect{4} defect is different from the lowest energy one found by
Hastings \etal\cite{Hastings:PhysicaB:1999} within HF theory, which
has a (1)-(1)-(1)-(1) configuration.
We have not been able to obtain the lowest energy \defect{4} structure from Ref. \onlinecite{Gharaibeh:PRB:2001}, although from the description given, we know that it is different from ours.
We are therefore unable to perform a full comparison of energies for \defect{4}.

\section{Discussion} 
\label{discussion}

We have presented first-principles DFT results using random structure
searching for hydrogen/silicon complexes in silicon.
The searches were carried out in 32-atom silicon cells, while the
final results were obtained with 256-atom cells.
We used a multi-k-point generalization of the Baldereschi mean-value
method to perform the BZ sampling, which we demonstrated to be
superior to the standard MP sampling.

Formation energies of the defects were calculated with respect to the
lowest energy self-interstitial defect and lowest energy hydrogen
defect in bulk silicon.
We have confirmed that the previously described \defect{2} and
\defect{2}$^{*}$ defects are the most stable.
We have, however, found a new \defect{} defect which is significantly
lower in energy than the one previously reported in the literature.
Our defect is based on the hexagonal self-interstitial whereas the
previously-reported one was based on the split-$\langle110\rangle$
self-interstitial.
We also found a new, lower energy, \defect{3} defect and a new \defect{4} defect.

\begin{figure}
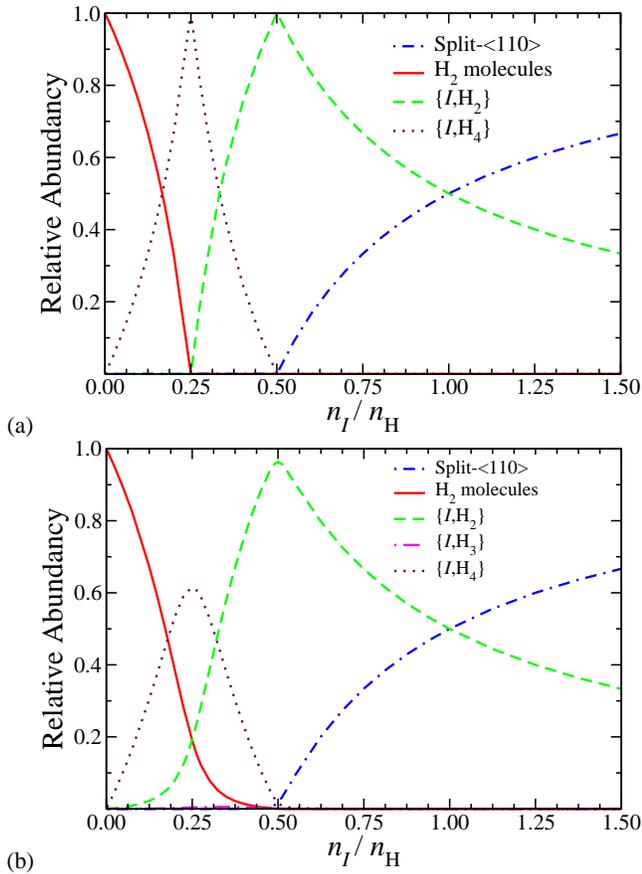

(a)
\includegraphics*[width=8cm]{fig8a.eps}\\
(b)
\includegraphics*[width=8cm]{fig8b.eps}
\caption[]{(Color online) Relative abundances of the defects at zero
temperature (a) and 1200~K (b). $n_I/n_{\mathrm{H}}$ is the ratio of the concentration of
interstitial silicon atoms to hydrogen atoms. At low
$n_I/n_{\mathrm{H}}$ there is a large relative abundance of H$_2$
molecules.  As $n_I/n_{\mathrm{H}}$ increases, H$_2$ molecules bind
strongly to the Si self-interstitials, forming \defect{4} defects.  As
$n_I/n_{\mathrm{H}}$ increases further, formation of \defect{2}
defects is favored.  However, an increase in $n_I/n_{\mathrm{H}}$
above $0.5$ does not lead to formation of \defect{}, because the mixed
state of $\left \{I \right \}$ and \defect{2} is more favorable.  The
most significant differences at 1200~K are that the abundance of
\defect{4} is somewhat reduced and that H$_2$ and \defect{2} defects
are favored instead, and \defect{3} has a small but finite abundance
in a region which peaks at around $n_I/n_{\mathrm{H}}=0.4$.  The
abundance of \defect{} at 1200~K is negligible.}
\label{fig:relative_abundances}
\end{figure}

The relative abundances of the defects at zero temperature can be
calculated from the defect energies as a function of the ratio of the
concentrations of the self-interstitial and hydrogen atoms,
$n_I/n_{\rm H}$.
Fig.~\ref{fig:relative_abundances}a shows that only four defects can
form in this model, $\{I\}$, \defect{2}, \defect{4}, and $\{{\rm
H}_2\}$, with the $\{I\}$ defect corresponding to the lowest-energy
split-$\langle110\rangle$ self-interstitial.
The main features of Fig.~\ref{fig:relative_abundances} are that when
$n_I \ll n_{\rm H}$ \defect{4} defects are formed and the surplus H
atoms form $\{{\rm H}_2\}$ defects, and when $n_I \gg n_{\rm H}$
\defect{2} defects are formed and the surplus Si atoms form $\{I\}$
defects.

At finite temperatures, see Fig.~\ref{fig:relative_abundances}b, we consider the defect free energies, which
should contain contributions from the vibrational free energy and the
configurational entropy.  We have not evaluated the vibrational free
energies, which would involve very costly phonon calculations, but we
have calculated the configurational contributions.  These are expected
to be significant because we deal with defects containing from one
atom (split-$\langle110\rangle$ self-interstitial) up to five atoms
(\defect{4}).  In general, defects containing fewer atoms are expected
to be favoured by the configurational entropy at higher temperatures.
The configurational entropy can be written in terms of the number of
degenerate defect configurations per atomic site, $d_i$, where $i$
labels the defect.  The degeneracy $d_i$ can be evaluated
straightforwardly in some cases.  For example, the \defect{} defect of
Fig.~\ref{Fig:I1H1} has a degeneracy per atomic site of four because
it is formed from a Si atom at a hexagonal site (of which there are
two per atomic site) and the Si-H bond can point in one of two
directions.  Calculating the degeneracy for more complicated defects
is not necessarily straightforward, and therefore we have developed a
computational scheme to evaluate defect degeneracies.  The scheme
comprises the following steps:
\begin{enumerate}
\item Generate a set of structures by
applying the symmetry operations of the space group of the host
crystal to the defect structure;
\item Identify structures from this
set which differ only by a translation vector of the lattice of the
host crystal and remove all but one of them;
\item The defect
degeneracy per primitive unit cell is the number of structures
remaining.
\end{enumerate} 
The defect degeneracies calculated in this fashion are
reported in Tables~\ref{Table:EF_H_Si} and \ref{Table:EF_H-Si_Si}.  Where 
we present $d_i$ only for the lowest energy defect of each type.
The defect concentrations are then obtained by minimizing the
Helmholtz free energy of the system for fixed concentrations $n_I$ and
$n_{\mathrm{H}}$.    In our model we consider only the lowest energy defects of each type
listed in Tables~\ref{Table:EF_H_Si} and \ref{Table:EF_H-Si_Si}, and
in a complete model other defects such as clusters of
self-interstitial Si atoms and complexes involving other impurity
atoms should be considered.

It is interesting to note that only the \defect{2}, \defect{2}$^*$,
and \defect{4} defects are perfectly saturated, \emph{i.e.}, each Si
atom has four covalent bonds and each H atom has one, and that these
defects have the lowest total formation energies (defined as $nE_{\rm
F}(\left\{ I, {\rm H}_n \right\})$ with $E_{\rm F}$ given by
Eq.~(\ref{Eqn:SiH-int})).
It is, of course, not possible to achieve perfect saturation of a Si/H
structure with an odd number of H atoms, as such a structure would
contain two bonds per Si atom and half a bond per H atom.

Overall we conclude that ``random searching'' is a useful tool for
finding the structures of low-energy defects in semiconductors.

\begin{acknowledgments}
We are grateful to Jonathan Yates, Michael Rutter, Phil Hasnip and
James Kermode for useful discussions.  This work was supported by the
Engineering and Physical Sciences Research Council (EPSRC) of the UK.
Computational resources were provided by the Cambridge High
Performance Computing Service.
\end{acknowledgments} 

\bibliographystyle{h-physrev}

\end{document}